%% file: paper_v12.tex
\RequirePackage{lineno}
\documentclass[twocolumn,aps,prl,showpacs,superscriptaddress,floatfix]{revtex4}
\usepackage{epsfig}
\usepackage{color}
\usepackage{lineno}
\newcommand{\snn}{\sqrt{s_{_\mathrm{NN}}}}
\newcommand{\gevc}{GeV/$c$}
\newcommand{\px}{P_x}
\newcommand{\pt}{p_T}
\newcommand{\kt}{k_T}
\newcommand{\deta}{\Delta\eta}
\newcommand{\dphi}{\Delta\phi}
\newcommand{\ptt}{p_T^{\rm trig}}
\newcommand{\pta}{p_T^{\rm assoc}}
\newcommand{\etat}{\eta_{\rm trig}}
\newcommand{\etaa}{\eta_{\rm assoc}}
\newcommand{\phit}{\phi_{\rm trig}}
\newcommand{\phia}{\phi_{\rm assoc}}
\newcommand{\Ntrig}{N_{\rm trig}}
\newcommand{\zvtx}{z_{vtx}}

\newcommand{\dsig}{\Delta\sigma}

\begin{document}
\title{Measurement of away-side broadening with self-subtraction of flow in Au+Au collisions at $\snn=200$~GeV}
%\author{STAR Collaboration}
\input author-20190619.tex

\date{\today}

\begin{abstract}
High transverse momentum ($\pt$) particle production is suppressed due to parton (jet) energy loss in the hot dense medium created in relativistic heavy-ion collisions. Redistribution of energy at low-to-modest $\pt$ has been elusive to measure because of large anisotropic backgrounds. 
We report a data-driven method for background evaluation and subtraction, exploiting the away-side pseudorapidity gaps, to measure the jetlike correlation shape in Au+Au collisions at $\snn=200$~GeV with the STAR experiment.
The correlation shapes, for trigger particle $\pt>3$~\gevc\ and various associated particle $\pt$ ranges within $0.5<\pt<10$~\gevc, are consistent with Gaussians and their widths are found to increase with centrality. 
The results indicate jet broadening in the medium created in central heavy-ion collisions.
\end{abstract}
%25.75.-q 	Relativistic heavy-ion collisions  
%25.75.Ag 	Global features
%25.75.Bh 	Hard scattering
%25.75.Cj 	Photon, lepton, and heavy quark production
%25.75.Dw 	Particle and resonance production
%25.75.Gz 	Particle correlations and fluctuations
%25.75.Ld 	Collective flow
%25.75.Nq 	Quark deconfinement, QGP production, and phase transitions (see also 12.38.Mh Quark-gluon plasma in quantum chromodynamics; 21.65.Qr Quark matter in nuclear matter)
\maketitle
%\linespread{1.6}

%%%%%%%%%%%%%%%%%%%%%%%%%%%%%%%%%%%%%%%%%%%%%%%%%%%%%%%%%%%%%%%
{\em Introduction.}
The basic constituents of nuclear matter are quarks and gluons. Their interactions are governed by quantum chromodynamics (QCD). QCD matter, normally confined into hadrons, is deconfined into a state of matter known as the quark-gluon plasma (QGP) under extreme conditions of high energy/matter densities~\cite{Shuryak:1978ij}. A QGP phase existed in the early universe and is also created in heavy-ion collisions at the Relativistic Heavy-Ion Collider (RHIC)~\cite{Arsene:2004fa,Back:2004je,Adcox:2004mh,Adams:2005dq} and the Large Hadron Collider (LHC)~\cite{Muller:2012zq}. One important piece of evidence for the discovery of the QGP is jet quenching, i.e.~parton (jet) energy loss in the QGP medium which in relativistic heavy-ion collisions results in suppression of high transverse momentum ($\pt$) particle and jet production~\cite{Adcox:2001jp,Adler:2002xw,Adler:2002tq,Adams:2003kv,Adler:2003ii,Adler:2003qi,Adams:2003im,Aad:2010bu,Chatrchyan:2011sx,Aamodt:2010jd}. The suppression is so strong that a density of at least 30 times normal nuclear density is required to describe data in model calculations~\cite{Jacobs:2004qv}.

The partonic energy loss mechanisms are, however, less clear. Some models focus on collisional and radiative energy losses~\cite{Jacobs:2004qv}. %Wang:1991xy,Gyulassy:1990ye,
Others propose more exotic mechanisms, such as collective excitation modes~\cite{Stoecker:2004qu,CasalderreySolana:2004qm,Ma:2010dv,Tachibana:2014lja,Tachibana:2015qxa,Tachibana:2017syd}. 
While single particle measurements are not sufficiently sensitive to energy loss mechanisms, measurements of how the lost energy is redistributed at low to modest $\pt$ are expected to be more sensitive. 
One way to this end is to reconstruct jets and study $\pt$ and angular distributions of jet fragments~\cite{Adare:2012qi,Khachatryan:2016erx,Chatrchyan:2013kwa}.
Distributions of the lost energy can also be measured via dihadron angular correlations with respect to high-$\pt$ trigger particles and jets. Previous measurements of two- and multi-particle correlations, after subtracting elliptic flow background, 
have revealed novel correlation structures~\cite{Adams:2005ph,Adler:2005ee,Adare:2008ae,Aggarwal:2010rf,Wang:2013qca}. However, due to initial collision geometry fluctuations, all orders of harmonics (not just elliptic) flow anisotropies are possible~\cite{Alver:2010gr,Heinz:2013th}. 
Full subtraction of anisotropic backgrounds is challenging and suffers from large uncertainties~\cite{Wang:2013qca,Agakishiev:2010ur,Agakishiev:2014ada,Adare:2018wjb,Aamodt:2011vg,Adam:2016xbp,STAR:2016jdz,Adamczyk:2013jei,Adamczyk:2012eoa,Agakishiev:2011st,Abelev:2016dqe,Adare:2012qi}. 

Here we devise a data-driven method with an ``automatic'' subtraction of anisotropic flow backgrounds. 
The method was first tested with toy model and PYTHIA simulations~\cite{Zhang:2019lmn}.
Although the correlated jetlike yield cannot be readily determined from this method, the correlation shape can be obtained without the large uncertainty from flow subtraction. We study the correlation shape as a function of the collision centrality and associated particle $\pt$. The correlation shape should be sensitive to the nature of jet-medium interactions, and therefore offers new opportunities to investigate energy loss mechanisms and medium properties.

{\em Experiment and Data.}
The data reported here were taken in 2011 by the STAR experiment using a minimum bias (MB) trigger in Au+Au collisions at the nucleon-nucleon center-of-mass energy of $\snn=200$~GeV. The MB trigger is defined by a coincidence signal between the east and west Vertex Position Detectors (VPD)~\cite{Llope:2003ti} located at the pseudorapidity range of $4.4<|\eta|<4.9$. A total of $3.2\times10^8$ MB trigger events are used. The event centrality is defined by the measured charged particle multiplicity within $|\eta|<0.5$. Data are reported in four centrality bins corresponding to 0-10\%, 10-30\%, 30-50\%, and 50-80\% of the total hadronic cross section~\cite{Abelev:2008ab}. 

The main detector used for this analysis is the Time Projection Chamber (TPC)~\cite{Ackermann:1999kc,Anderson:2003ur}, residing in a 0.5~T magnetic field along the beam direction ($z$). 
Particle tracks are reconstructed in the TPC and are required to have at least 20 out of 45 maximum possible hits. 
%Track splitting is eliminated via the technique outlined in~\cite{Abelev:2008ab}.
The ratio of the number of hits used in track reconstruction to the number of possible hits is required to be greater than 0.51 to eliminate multiple track segments being reconstructed from a single particle trajectory.
The primary vertex (PV) is reconstructed using tracks. Events with a PV position ($\zvtx$) within $30$ cm of the TPC center along $z$ are used. To remove secondaries from particle decays, only tracks with the distance of closest approach DCA $<2$ cm to the PV are used. 

{\em Analysis Method.}
Jetlike correlations are studied with respect to high-$\pt$ trigger particles, which serve as proxies for jets~\cite{Jacobs:2004qv,Wang:2013qca}. 
High-$\pt$ particles measured at RHIC are strongly biased toward the surface of the collision zone~\cite{Jacobs:2004qv,Zhang:2007ja,Renk:2012ve}. The away-side jet partner that is preferentially directed inward, is therefore very likely to traverse the entire volume suffering maximal interactions with the medium. Because of the broad distribution of the underlying parton kinematics, the away-side jet direction is mostly uncorrelated in $\eta$ relative to the trigger particle~\cite{Wang:2013qca}. 
It is therefore difficult to distinguish the jet signal from the underlying background; the large, azimuthally anisotropic background has to be {\em specifically} subtracted, with large uncertainties, traditionally using measured anisotropy parameters~\cite{Wang:2013qca}. 
The away-side jet direction can be localized by requiring a second high-$\pt$ particle back-to-back in azimuthal angle ($\phi$) with respect to the first one. However, by doing so, the back-to-back dijets are biased towards being tangential to the collision zone~\cite{Agakishiev:2011nb,Adamczyk:2012eoa}, substantially weakening the purpose of studying jet-medium interactions. 

\begin{figure}[thb]
  \centering
  \includegraphics[width=\columnwidth]{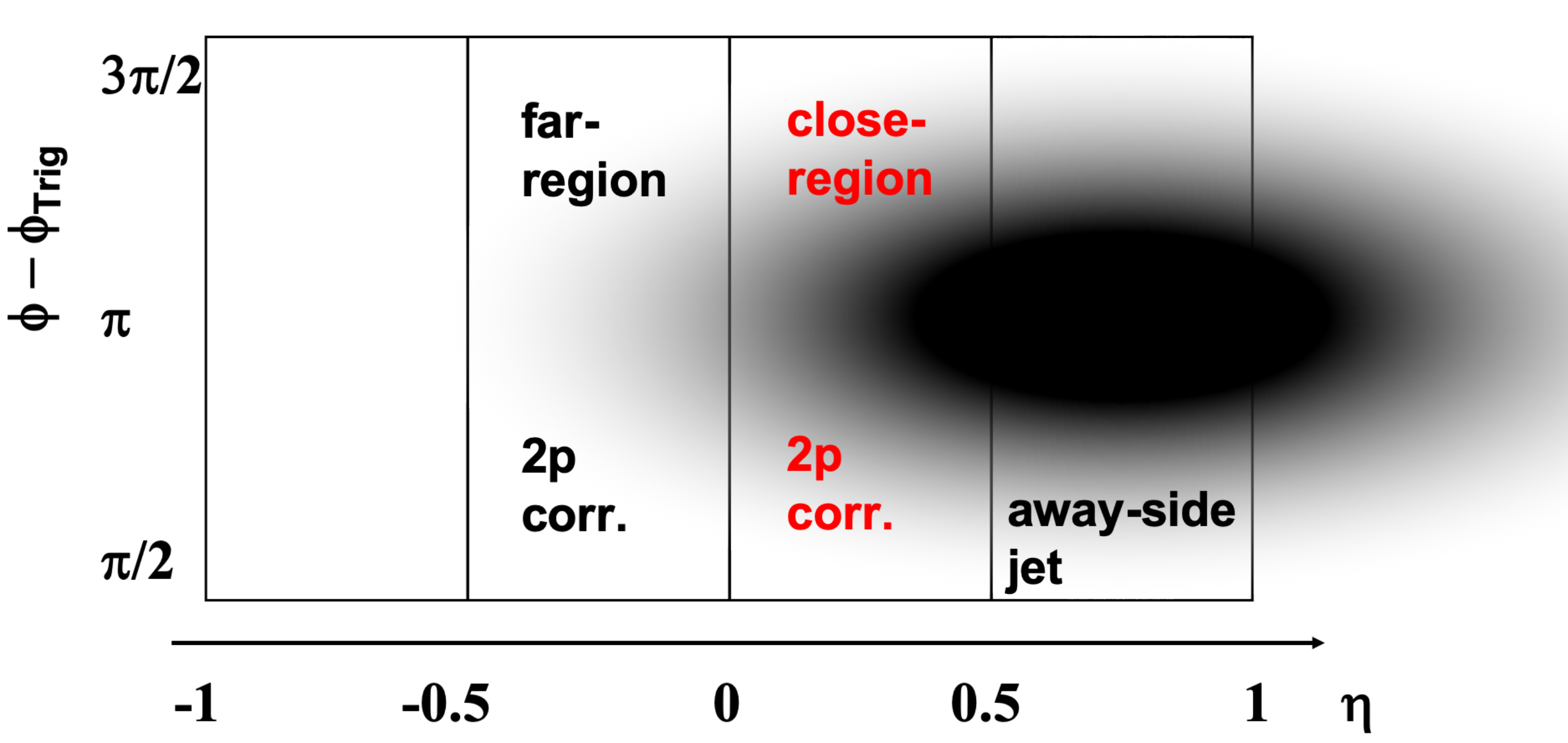}
  \caption{Schematic diagram of the analysis method. The requirement of a large recoil $\px$ [Eq.~(\ref{eq:px})] in a particular $\eta$ region ($0.5<\eta<1$ shown here) selects events with enhanced population of jets close to the $\eta$ region. Jetlike correlations in the close-region and far-region, symmetric about midrapidity, contain different contributions from the jet but the same contribution from the flow background. Their difference measures the jetlike correlation shape.}
  \label{fig:cartoon}
\end{figure}

In this analysis, we impose a less biasing requirement of a large recoil transverse momentum ($\px$) azimuthally opposite to the high-$\pt$ trigger particle, within a given $\eta$ range, to enhance the away-side jet population in the acceptance. The schematic diagram in Fig.~\ref{fig:cartoon} shows the away-side $\eta$-$\phi$ space and illustrates by the fading gray area the away-side jet population enhanced in a particular $\eta$ regoin.
$\px$ is given by
\begin{equation}
\px|_{\eta_1}^{\eta_2}=\sum_{\eta_1<\eta<\eta_2,|\phi-\phit|>\pi/2}\pt\cos(\phi-\phit)\cdot\frac{1}{\epsilon}\;,
\label{eq:px}
\end{equation}
where all charged particles ($0.15<\pt<10$~\gevc) within the $\eta$ range that are on the away side ($|\phi-\phit|>\pi/2$) of the trigger particle are included. Since the near-side jet is not included in the $P_{x}$ calculation, the $\eta$ distribution of the trigger particle is unbiased by the $\px$ cut. 
The inverse of the single-particle relative acceptance$\times$efficiency ($\epsilon$) is used to correct for the single-particle detection efficiency. It depends on the position of the primary vertex along the beam axis $\zvtx$, collision centrality, particle $\pt$, $\eta$ and $\phi$, and has run period variations~\cite{Abelev:2008ab}. 
The $\phi$-dependence of $\epsilon_{\phi}$ is obtained, separately for positive and negative $\eta$, from the single-particle $\phi$ distribution normalized to unity on average in each centrality. 
The $\eta$-dependence of $\epsilon$ varies with $\zvtx$, centrality and $\pt$, and is obtained by treating symmetrized $dN/d\eta$ distribution in events with $|\zvtx|<2$~cm as the baseline, and taking the ratio of the $dN/d\eta$ distribution from each $\zvtx$ bin to this baseline. 
Because our $\px$ cut is only used to select a given fraction of events, the absolute efficiency correction is not applied. 

\begin{figure*}[thb]
  \centering
  \includegraphics[width=0.7\textwidth]{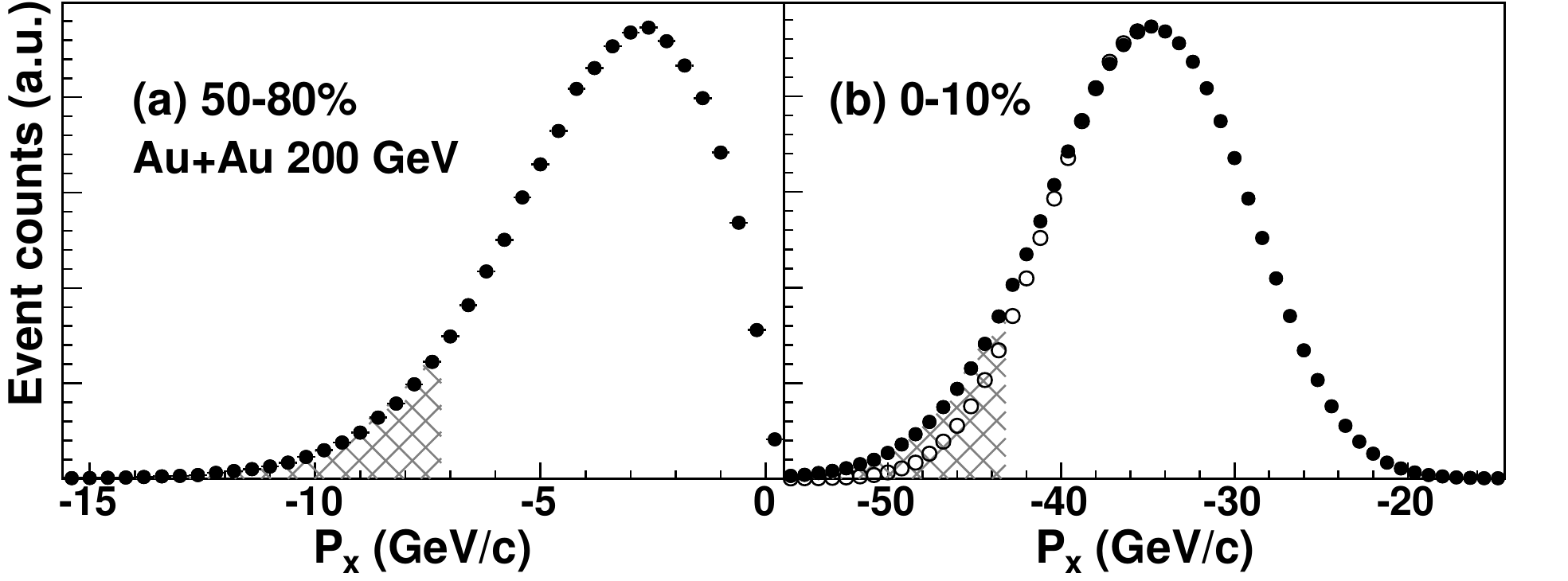}
  \caption{Distributions of the recoil momentum within $0.5<\eta<1$ ($\px|_{0.5}^{1}$) from high-$\pt$~trigger particles of $3<\ptt<10$~\gevc\ in (a) 50-80\% peripheral and (b) 0-10\% central collisions. The shaded areas indicate a selection of 10\% of events to enhance the away-side jet population inside acceptance. Events from $|\zvtx|<2$~cm are used; other $\zvtx$ events and $\px|_{-1}^{-0.5}$ are similar. The right side of the $P_x$ distribution is reflected to the left of the maximum as the open circles.}
  \label{fig:px}
\end{figure*}

In this analysis the trigger particle $\pt$ range is $3<\ptt<10$~\gevc. We choose the windows $-1<\eta<-0.5$ or $0.5<\eta<1$ for $\px$ calculation. 
Figure~\ref{fig:px} shows example $\px|_{0.5}^{1}$ distributions for peripheral and central Au+Au collisions. Their difference comes mainly from event multiplicities. For each centrality, we select the 10\% of the events with the highest $-\px$ to enhance the probability that the away-side jet population is contained in this $\eta$ region. There is a large statistical fluctuation effect in $\px$, especially in central collisions. The $\px$ selection may also be affected by low-$\pt$ minijets.
These effects do not strictly give a symmetric distribution~\cite{Zhang:2019lmn}. Nevertheless, we show the reflected data as the open circles in Fig.~\ref{fig:px}(b) to give an order of magnitude estimate of those effects.

In the selected events, we analyze dihadron correlations of associated particles, with respect to trigger particles, in two $\eta$ regions symmetric about midrapidity, one close (``close-region'') to and the other far (``far-region'') from the $\eta$ window for $\px$. See the sketch in Fig.~\ref{fig:cartoon}.
The dihadron correlation in $\dphi=\phia-\phit$, between the associated and trigger particle azimuthal angles, is given by
\begin{equation}
  \frac{dN}{d\dphi}=\frac{1}{\Ntrig}\cdot\frac{S(\dphi)}{B(\dphi)/B_0}\;,
  \label{eq:corr}
\end{equation}
where
\begin{equation}
  S(\dphi)=\int_{-1}^{+1}d\etat\int_{\rm region}d\etaa\frac{d^3N}{d\etat d\etaa d\dphi}\cdot\frac{1}{\epsilon}\label{eq:S}
\end{equation}
and $B(\dphi)$ is its counterpart from mixed events. The correlations are normalized by the number of trigger particles, $\Ntrig$. In Eq.~(\ref{eq:S}), ``region'' stands for close-region or far-region. All the trigger particles with $|\etat|<1$ are integrated. 
The single-particle relative acceptance$\times$efficiency ($\epsilon$) correction is applied for associated particles.
Like in $\px$, the absolute efficiency correction is not applied in the correlation measurements because this analysis deals with only the correlation shape, not the absolute amplitude. 
The mixed-events are formed by pairing the trigger particles in each event with the associated particles from 10 different random events in the same centrality and $\zvtx$ bin.
The mixed-event background $B(\dphi)$ is normalized to unity (via the constant $B_0$) to correct for residual two-particle acceptance after single particle efficiency correction. 

The away-side jet contributes more to the close region than to the far region due to the larger $\deta$ gap of the latter (see the sketch in Fig.~\ref{fig:cartoon}). The anisotropic flow contributions, on the other hand, are on average equal in these two regions that are symmetric about midrapidity. The difference in the close- and far-region correlations, therefore, arises only from jetlike correlations. 
For $\px|_{-1}^{-0.5}$, the close-region is $-0.5<\eta<0$ and the far-region is $0<\eta<0.5$; for $\px|_{0.5}^{1}$, they are swapped. 
The results from these two sets are consistent, and thus combined. 
We exclude events where both $\px|_{-1}^{-0.5}$ and $\px|_{0.5}^{1}$ satisfy the respective 10\% $P_x$ cut, because the combined signal would be strictly zero but with a propagated nonzero statistical error.

PYTHIA simulations~\cite{Zhang:2019lmn} indicate that jet fragmentations are approximately factorized in $\eta$ and $\phi$. The $\dphi$ correlations at different $\deta$ have approximately the same shape, only differing in magnitude. Thus, the difference between close- and far-region correlations measures the away-side correlation shape. We quantify the shape by Gaussian width $\sigma$ determined from a fit. 

{\em Systematic Uncertainties.}
The systematic uncertainties of $\sigma$ come from several sources. Varying the $\px$ cut changes the relative contributions of jets and background fluctuations to the selected events, but should not affect the correlation width significantly if the jet sample is unbiased. We vary the $\px$ cut from allowing the default $10\%$ of events to 2\%, 5\%, 15\%, 20\%, 30\% and 50\% of events. The calculated systematic uncertainty of $\sigma$ is 3.4\% (one standard deviation).

We have assumed that jetlike correlations are factorized in $\eta$ and $\phi$. There is theoretical~\cite{Bozek:2010vz,Xiao:2012uw,Jia:2014ysa} and experimental evidence~\cite{Khachatryan:2015oea} that flow may be decorrelated over $\eta$ due to geometry fluctuations. Both these effects would cause uncertainties in attributing the close- and far-region difference purely to jetlike correlations. We vary the close- and far-region $\eta$ locations and ranges so they have different $\eta$ gaps in between as well as from the $\px$ $\eta$ window, but still symmetric about midrapidity. We also vary the $\px$ $\eta$ window location and range. The largest deviation of $\sigma$ from the default results is approximately half of the statistical error. The calculated systematic uncertainty of $\sigma$ is 2.0\% (one standard deviation) for this source.

In addition, we vary the track quality cuts in the analysis. The calculated systematic uncertainty for this source is 5.3\% standard deviation in $\sigma$. The final systematic uncertainty is calculated as the quadratic sum of all the sources we studied.

The systematic uncertainties on $\sigma$ are found to be partially correlated between various centralities and $\pta$ bins. In the difference between central and peripheral collisions, $\dsig=\sqrt{\sigma^2_{\rm cent}-\sigma^2_{\rm peri}}$, the systematic uncertainties are not simply propagated but obtained in the same way as those on the individual $\sigma$'s described above. The same procedure is used to obtain the systematic uncertainty on the linear parameterization of $\dsig$ versus $\pta$.

\begin{figure}[thb]
  \centering
  \includegraphics[width=0.8\columnwidth]{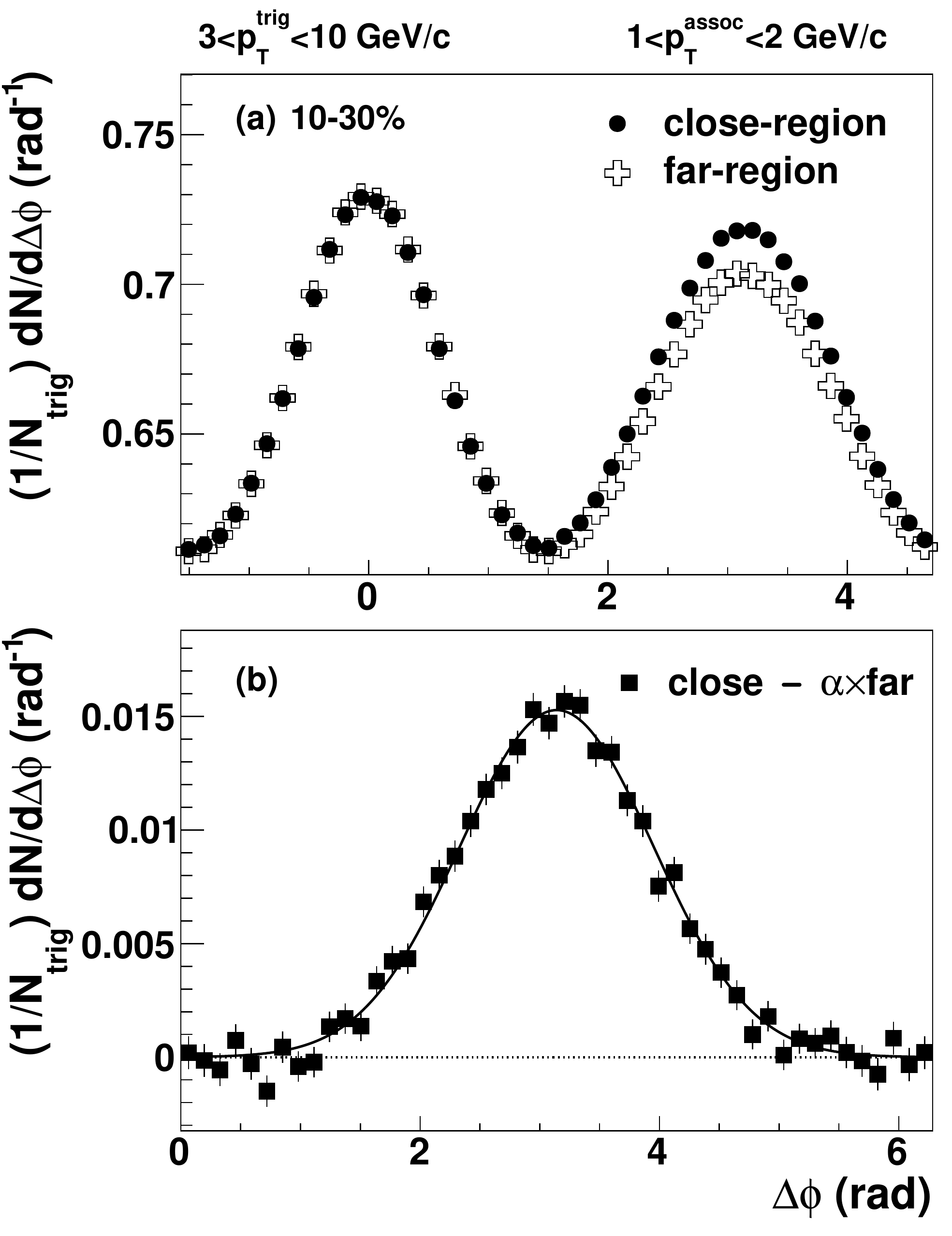}
  \caption{(a) Dihadron azimuthal correlations in close-region (solid circles) and far-region (open crosses), as an example, for $3<\ptt<10$~\gevc\ and $1<\pta<2$~\gevc\ in 10-30\% Au+Au collisions at $\snn=200$~GeV. (b) The difference between close-region correlation and scaled far-region correlation (see text for detail). The curve is a Gaussian fit with the centroid fixed at $\pi$. Errors are statistical.}
  \label{fig:corr}
\end{figure}

{\em Results and Discussions.}
Figure~\ref{fig:corr}(a) shows, as an example, the dihadron azimuthal correlations for the close-region and far-region in 10-30\% Au+Au collisions at $\snn=200$~GeV for trigger $3<\ptt<10$~\gevc\ and associated particle $1<\pta<2$~\gevc.
The near-side correlations are almost identical for close- and far-region. The near-side ratio of far- to close-region correlations, $\alpha$, are listed in Table~\ref{tab:alpha} and are all approximately unity. 
This indicates, to a good degree, that the near-side jetlike correlations are not biased by the $\px$ selection and flow contributions to close- and far-region correlations are indeed equal.
\begin{table*}
  \caption{The near-side ratio of far- to close-region correlations, $\alpha$, averaged over $|\dphi|<1$ as a function of $\pta$ and centrality in Au+Au collisions at $\snn=200$~GeV. The trigger particle has $3<\ptt<10$~\gevc. Errors are statistical.}
  \label{tab:alpha}
  \begin{tabular}{c|c|c|c|c}
    $\pta$ ($\gevc$) &50-80\% & 30-50\% & 10-30\% & 0-10\% \\\hline
    0.15-0.5 & $1.0038\pm0.0007$ & $1.0036\pm0.0003$ & $1.0027\pm0.0001$ & $1.0002\pm0.0001$ \\
    0.5-1 & $1.000\pm0.001$ & $1.0021\pm0.0003$ & $1.0006\pm0.0002$ & $0.9984\pm0.0001$ \\
    1-2 & $1.002\pm0.002$ & $1.0002\pm0.0006$ & $0.9994\pm0.0003$ & $0.9976\pm0.0002$ \\
    2-3 & $0.997\pm0.006$ & $1.006\pm0.002$ & $0.9995\pm0.0009$ & $0.9965\pm0.0008$ \\
    3-10 & $0.99\pm0.01$ & $0.999\pm0.005$ & $0.998\pm0.003$ & $0.999\pm0.003$
\end{tabular}
\end{table*}

The away-side correlations differ in amplitude and shape which is caused by the away-side jet contributions. The far-region correlation is scaled by $\alpha$ to account for the small near-side difference and then we subtract it from the close-region correlation. 
The result is shown in Fig.~\ref{fig:corr}(b). The difference measures the away-side jetlike correlation shape. A Gaussian fit centered at $\Delta\phi=\pi$ is applied to extract the correlation width. The $\chi^2$ values per degree of freedom are all consistent with unity, indicating that the correlation shape is Gaussian. The $\sigma$ values are tabulated in Table~\ref{tab:sigma}. 
\begin{table*}
  \caption{Gaussian fit width to away-side jetlike correlations as a function of $\pta$ and centrality in Au+Au collisions at $\snn=200$~GeV. The trigger particle has $3<\ptt<10$~\gevc. The first error is statistical and the second is systematic.}
  \label{tab:sigma}
  \begin{tabular}{c|c|c|c|c}
    $\pta$ ($\gevc$) &50-80\% & 30-50\% & 10-30\% & 0-10\% \\\hline
    0.15-0.5 & $0.98\pm0.04\pm0.06$ & $0.96\pm0.03\pm0.06$ & $1.07\pm0.03\pm0.07$ & $0.99\pm0.03\pm0.06$ \\
    0.5-1 & $0.87\pm0.02\pm0.06$ & $0.84\pm0.02\pm0.06$ & $0.91\pm0.02\pm0.06$ & $0.94\pm0.03\pm0.06$ \\
    1-2 & $0.72\pm0.02\pm0.05$ & $0.79\pm0.02\pm0.05$ & $0.81\pm0.02\pm0.05$ & $0.83\pm0.02\pm0.06$ \\
    2-3 & $0.56\pm0.03\pm0.04$ & $0.67\pm0.03\pm0.04$ & $0.75\pm0.03\pm0.05$ & $0.77\pm0.04\pm0.05$\\
    3-10 & $0.42\pm0.04\pm0.03$ & $0.59\pm0.05\pm0.04$ & $0.67\pm0.05\pm0.04$ & $0.82\pm0.09\pm0.05$
\end{tabular}
\end{table*}

\begin{figure}[thb]
\centering
\includegraphics[width=0.8\columnwidth]{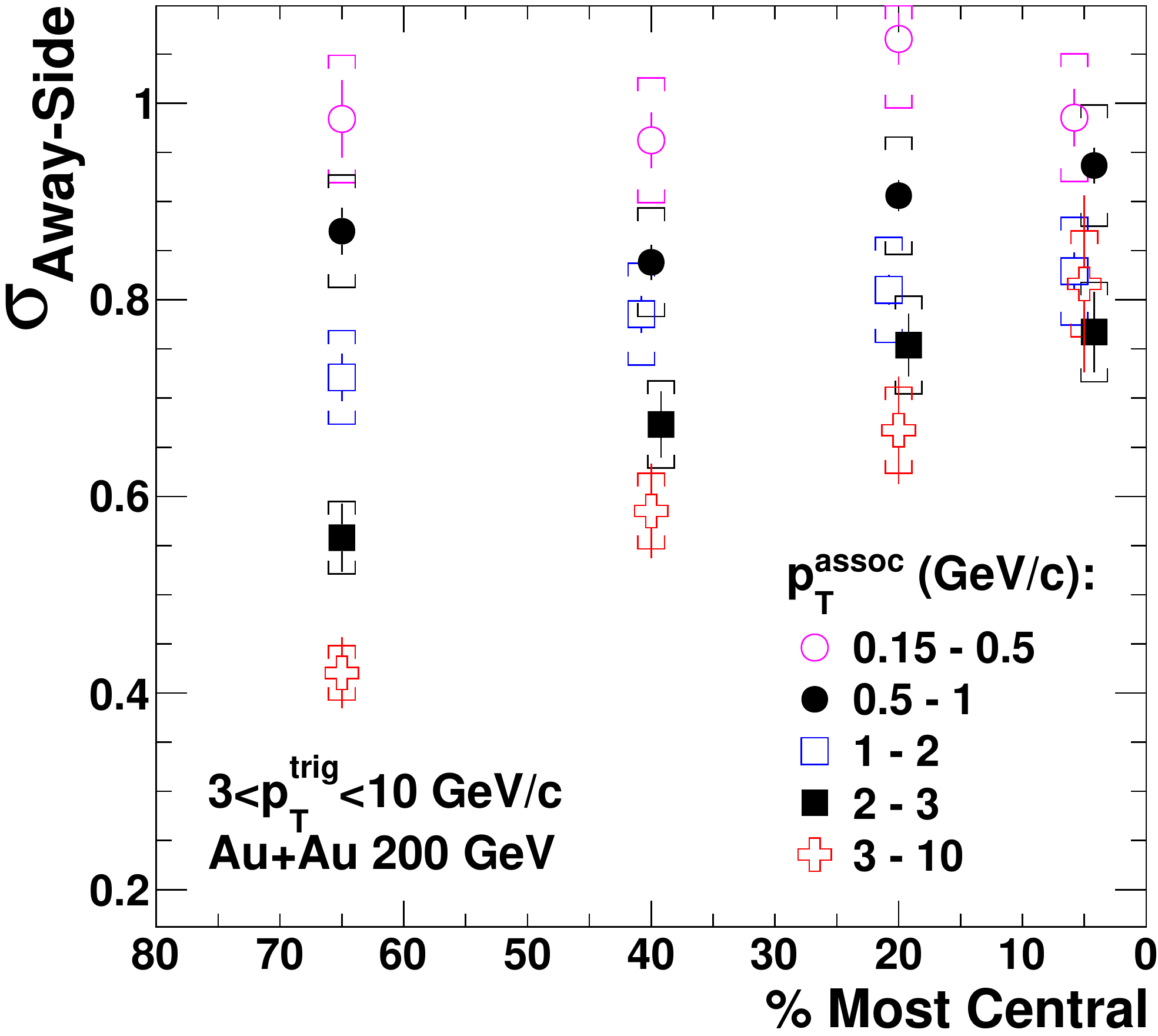}
\caption{Away-side jetlike correlation width (Gaussian $\sigma$) as a function of centrality (0 indicates the most central collisions) for $3<\ptt<10$~\gevc\ and various $\pta$ bins in Au+Au collisions at $\snn=200$~GeV. Bars are statistical errors and caps are systematic uncertainties.}
\label{fig:width}
\end{figure}

Figure~\ref{fig:width} shows the away-side correlation width (Gaussian $\sigma$) as a function of centrality for five $\pta$ bins. The width for the lowest $\pta$ of 0.15-0.5~\gevc\ is consistent with a constant over centrality; at this low $\pta$, the correlations are fairly wide for all centralities and possible broadening with increasing centrality may not be easily observable. For the four higher $\pta$ bins, the width increases from peripheral to central collisions. 
The broadening of the correlation function is consistent with jet broadening. However, it is also possible, because the correlation measurement is statistical, that the broadening comes from an increasing dijet acoplanarity (nuclear $\kt$ effect~\cite{Cronin:1974zm,Vitev:2004kd}) with increasing centrality. 
One possible mechanism of nuclear $\kt$ effect is multiple scattering by the incident nucleons before the hard scattering between underlying partons. 
Nuclear $\kt$ effect gives the hard scattering partons an initial net momentum in the transverse plane and thus causes a spread in dijet angular correlation. 

\begin{figure}[thb]
\centering
\includegraphics[width=0.8\columnwidth]{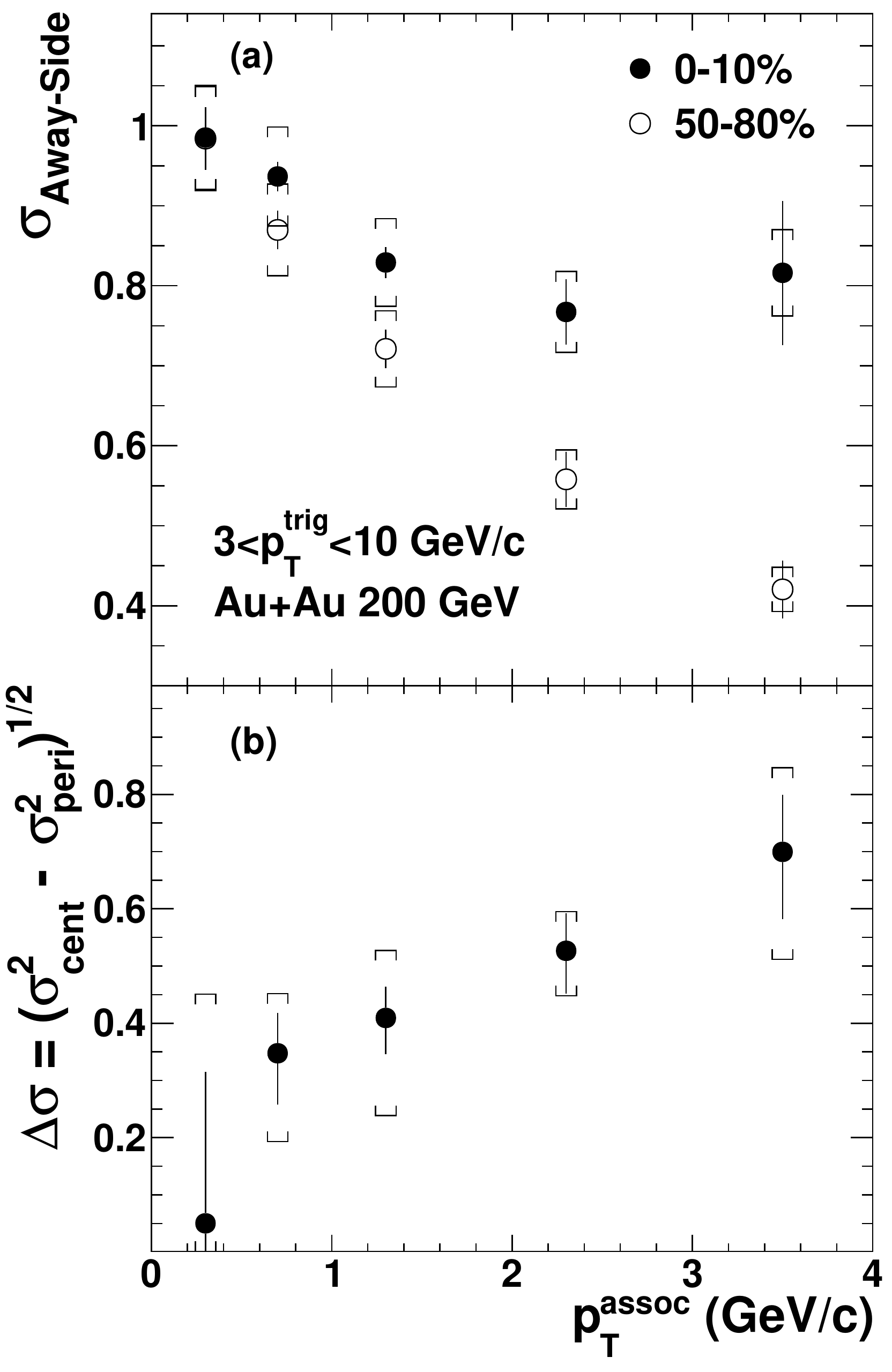}
\caption{(a) Away-side jetlike correlation width as a function of $\pta$ for $3<\ptt<10$~\gevc\ in 50-80\% peripheral (open circles) and top 10\% central (solid circles) Au+Au collisions at $\snn=200$~GeV. (b) The difference between the correlation widths in central and peripheral collisions. Bars are statistical errors and caps are systematic uncertainties.}
\label{fig:width_pt}
\end{figure}

Figure~\ref{fig:width_pt}(a) shows $\sigma$ as a function of $\pta$ in peripheral and central collisions. In peripheral collisions, the width decreases rapidly with increasing $\pta$. In central collisions the decrease is less rapid. We quantify the broadening from peripheral to central collisions by
$\dsig=\sqrt{\sigma^2_{\rm cent}-\sigma^2_{\rm peri}}$, shown as a function of $\pta$ in Fig.~\ref{fig:width_pt}(b). The relative broadening is stronger for higher $\pt$ associated particles. At very low $\pta$ the jetlike correlation is already quite broad in peripheral collisions, limiting any further broadening in central collisions. At high $\pta$ the initial jetlike correlation is narrow, leaving significant room for broadening in central collisions. 
In previous STAR dihadron correlation measurements~\cite{Adams:2005dq,Aggarwal:2010rf}, the reported away-side correlations were broader than those reported here likely because the previous results did not have the high-order harmonic flow backgrounds subtracted.
We also note that the reported jet-hadron correlations~\cite{Adamczyk:2013jei} were measured with a much higher jet $\pt$ and the extracted widths at low $\pta$ suffer from large flow background uncertainties.

If the away-side correlation broadening is due to nuclear $\kt$ effects only, without medium induced jet broadening, then we would have $\sigma^2_{\rm cent}=\sigma^2_{\rm peri}+\sigma^2_{\kt}$. Here $\kt$ quantifies the dijet acoplanarity and should ideally not depend on the associated particle $\pta$.
With the wide $\ptt$ range, it is possible that a higher $\pta$ could bias towards higher $\ptt$, hence smaller $\kt$ effect. To investigate this quantitatively, we fit the data in Fig.~\ref{fig:width_pt}(b) by a linear function, yielding
$\dsig=(0.23\pm0.09{\rm (stat.)}\pm0.11{\rm (syst.)})+(0.13\pm0.04{\rm (stat.)}\pm0.05{\rm (syst.)})\pt$ ($\pt$ in \gevc).
This suggests that the nuclear $\kt$ effect (expected constant or decreasing with $\pta$) is not the only source for the observed broadening. There must be contributions from a $\pt$ dependent effect such as medium induced jet broadening.
This conclusion is corroborated by the relatively small nuclear $\kt$ measured by both PHENIX~\cite{Rak:2004gk,Adler:2006sc} and STAR~\cite{Henry:2005wr,Putschke:2009wr}
We note that the measured broadening is between the associated and trigger particle angles, not directly the angle of jet fragment from the jet axis. It is the combination of broadening at the trigger particle $\pt$ and the associated particle $\pt$ values.

A more explicit means to distinguish the jet-medium broadening from the $\kt$ effect and other possible mechanisms is to use three-particle correlations~\cite{Abelev:2008ac,Abelev:2009jv}. With our method of subtracting anisotropic flow background, three-particle correlations could shed new light on partonic energy loss mechanisms in relativistic heavy-ion collisions. We leave such studies to future investigations.

{\em Conclusions.}
We have reported a measurement of away-side jetlike azimuthal correlation shapes relative to a high-$\pt$ trigger particle ($3<\ptt<10$~\gevc) in Au+Au collisions at $\snn=200$~GeV by the STAR experiment. We devised a method for a clean and robust subtraction of anisotropic flow backgrounds by using the correlation data itself. Namely, we enhance the away-side jet population in the acceptance by requiring a large recoil momentum $\px$ [see Eq.~(\ref{eq:px})], and take the difference of jetlike correlations in regions symmetric about midrapidity but with different $\deta$ gaps away from the enhanced $\px$ region. The measured Gaussian width of the away-side jetlike correlation increases with increasing centrality in the associated particle $\pt$ range of $0.5<\pta<10$~\gevc. The increase is consistent with medium induced jet broadening of the trigger and/or associated particles in addition to the nuclear $\kt$ effects. 

%%%%%%%%%%%%%%%%%%%%%%%%%%%%%%%%%%%%%%%%%%%%%%%%%%%%%%%%%%%%%%%
{\em Acknowledgments.}
We thank the RHIC Operations Group and RCF at BNL, the NERSC Center at LBNL, and the Open Science Grid consortium for providing resources and support.  This work was supported in part by the Office of Nuclear Physics within the U.S.~DOE Office of Science, the U.S.~National Science Foundation, the Ministry of Education and Science of the Russian Federation, National Natural Science Foundation of China, Chinese Academy of Science, the Ministry of Science and Technology of China and the Chinese Ministry of Education, the National Research Foundation of Korea, Czech Science Foundation and Ministry of Education, Youth and Sports of the Czech Republic, Hungarian National Research, Development and Innovation Office (FK-123824), New National Excellency Programme of the Hungarian Ministry of Human Capacities (UNKP-18-4), Department of Atomic Energy and Department of Science and Technology of the Government of India, the National Science Centre of Poland, the Ministry  of Science, Education and Sports of the Republic of Croatia, RosAtom of Russia and German Bundesministerium fur Bildung, Wissenschaft, Forschung and Technologie (BMBF) and the Helmholtz Association.

%%%%%%%%%%%%%%%%%%%%%%%%%%%%%%%%%%%%%%%%%%%%%%%%%%%%%%%%%%%%%%%
%\bibliographystyle{unsrt}
%\bibliography{iopart-num}
%\bibliography{../../rc/ref}
\input paper_v12.bbl

\end{document}

%% file: author-20190619.tex
\affiliation{Abilene Christian University, Abilene, Texas   79699}
\affiliation{AGH University of Science and Technology, FPACS, Cracow 30-059, Poland}
\affiliation{Alikhanov Institute for Theoretical and Experimental Physics, Moscow 117218, Russia}
\affiliation{Argonne National Laboratory, Argonne, Illinois 60439}
\affiliation{American Univerisity of Cairo, Cairo, Egypt}
\affiliation{Brookhaven National Laboratory, Upton, New York 11973}
\affiliation{University of California, Berkeley, California 94720}
\affiliation{University of California, Davis, California 95616}
\affiliation{University of California, Los Angeles, California 90095}
\affiliation{University of California, Riverside, California 92521}
\affiliation{Central China Normal University, Wuhan, Hubei 430079 }
\affiliation{University of Illinois at Chicago, Chicago, Illinois 60607}
\affiliation{Creighton University, Omaha, Nebraska 68178}
\affiliation{Czech Technical University in Prague, FNSPE, Prague 115 19, Czech Republic}
\affiliation{Technische Universit\"at Darmstadt, Darmstadt 64289, Germany}
\affiliation{E\"otv\"os Lor\'and University, Budapest, Hungary H-1117}
\affiliation{Frankfurt Institute for Advanced Studies FIAS, Frankfurt 60438, Germany}
\affiliation{Fudan University, Shanghai, 200433 }
\affiliation{University of Heidelberg, Heidelberg 69120, Germany }
\affiliation{University of Houston, Houston, Texas 77204}
\affiliation{Huzhou University, Huzhou, Zhejiang  313000}
\affiliation{Indian Institute of Science Education and Research (IISER), Berhampur 760010 , India}
\affiliation{Indian Institute of Science Education and Research, Tirupati 517507, India}
\affiliation{Indian Institute Technology, Patna, Bihar, India}
\affiliation{Indiana University, Bloomington, Indiana 47408}
\affiliation{Institute of Physics, Bhubaneswar 751005, India}
\affiliation{University of Jammu, Jammu 180001, India}
\affiliation{Joint Institute for Nuclear Research, Dubna 141 980, Russia}
\affiliation{Kent State University, Kent, Ohio 44242}
\affiliation{University of Kentucky, Lexington, Kentucky 40506-0055}
\affiliation{Lawrence Berkeley National Laboratory, Berkeley, California 94720}
\affiliation{Lehigh University, Bethlehem, Pennsylvania 18015}
\affiliation{Max-Planck-Institut f\"ur Physik, Munich 80805, Germany}
\affiliation{Michigan State University, East Lansing, Michigan 48824}
\affiliation{National Research Nuclear University MEPhI, Moscow 115409, Russia}
\affiliation{National Institute of Science Education and Research, HBNI, Jatni 752050, India}
\affiliation{National Cheng Kung University, Tainan 70101 }
\affiliation{Nile University, ECPT, 12677 Giza, Egypt}
\affiliation{Nuclear Physics Institute of the CAS, Rez 250 68, Czech Republic}
\affiliation{Ohio State University, Columbus, Ohio 43210}
\affiliation{Panjab University, Chandigarh 160014, India}
\affiliation{Pennsylvania State University, University Park, Pennsylvania 16802}
\affiliation{NRC "Kurchatov Institute", Institute of High Energy Physics, Protvino 142281, Russia}
\affiliation{Purdue University, West Lafayette, Indiana 47907}
\affiliation{Rice University, Houston, Texas 77251}
\affiliation{Rutgers University, Piscataway, New Jersey 08854}
\affiliation{Universidade de S\~ao Paulo, S\~ao Paulo, Brazil 05314-970}
\affiliation{University of Science and Technology of China, Hefei, Anhui 230026}
\affiliation{Shandong University, Qingdao, Shandong 266237}
\affiliation{Shanghai Institute of Applied Physics, Chinese Academy of Sciences, Shanghai 201800}
\affiliation{Southern Connecticut State University, New Haven, Connecticut 06515}
\affiliation{State University of New York, Stony Brook, New York 11794}
\affiliation{Temple University, Philadelphia, Pennsylvania 19122}
\affiliation{Texas A\&M University, College Station, Texas 77843}
\affiliation{University of Texas, Austin, Texas 78712}
\affiliation{Tsinghua University, Beijing 100084}
\affiliation{University of Tsukuba, Tsukuba, Ibaraki 305-8571, Japan}
\affiliation{United States Naval Academy, Annapolis, Maryland 21402}
\affiliation{Valparaiso University, Valparaiso, Indiana 46383}
\affiliation{Variable Energy Cyclotron Centre, Kolkata 700064, India}
\affiliation{Warsaw University of Technology, Warsaw 00-661, Poland}
\affiliation{Wayne State University, Detroit, Michigan 48201}
\affiliation{Yale University, New Haven, Connecticut 06520}

\author{J.~Adam}\affiliation{Brookhaven National Laboratory, Upton, New York 11973}
\author{L.~Adamczyk}\affiliation{AGH University of Science and Technology, FPACS, Cracow 30-059, Poland}
\author{J.~R.~Adams}\affiliation{Ohio State University, Columbus, Ohio 43210}
\author{J.~K.~Adkins}\affiliation{University of Kentucky, Lexington, Kentucky 40506-0055}
\author{G.~Agakishiev}\affiliation{Joint Institute for Nuclear Research, Dubna 141 980, Russia}
\author{M.~M.~Aggarwal}\affiliation{Panjab University, Chandigarh 160014, India}
\author{Z.~Ahammed}\affiliation{Variable Energy Cyclotron Centre, Kolkata 700064, India}
\author{I.~Alekseev}\affiliation{Alikhanov Institute for Theoretical and Experimental Physics, Moscow 117218, Russia}\affiliation{National Research Nuclear University MEPhI, Moscow 115409, Russia}
\author{D.~M.~Anderson}\affiliation{Texas A\&M University, College Station, Texas 77843}
\author{R.~Aoyama}\affiliation{University of Tsukuba, Tsukuba, Ibaraki 305-8571, Japan}
% ...

\collaboration{STAR Collaboration}\noaffiliation

%% file: paper_v12.bbl
\newcommand{\sNN}{$\sqrt{s_{NN}}$}

%% file: paper_v12.bbl
\begin{thebibliography}{63}
\expandafter\ifx\csname natexlab\endcsname\relax\def\natexlab#1{#1}\fi
\expandafter\ifx\csname bibnamefont\endcsname\relax
  \def\bibnamefont#1{#1}\fi
\expandafter\ifx\csname bibfnamefont\endcsname\relax
  \def\bibfnamefont#1{#1}\fi
\expandafter\ifx\csname citenamefont\endcsname\relax
  \def\citenamefont#1{#1}\fi
\expandafter\ifx\csname url\endcsname\relax
  \def\url#1{\texttt{#1}}\fi
\expandafter\ifx\csname urlprefix\endcsname\relax\def\urlprefix{URL }\fi
\providecommand{\bibinfo}[2]{#2}
\providecommand{\eprint}[2][]{\url{#2}}

\bibitem[{\citenamefont{Shuryak}(1978)}]{Shuryak:1978ij}
\bibinfo{author}{\bibfnamefont{E.~V.} \bibnamefont{Shuryak}},
  \bibinfo{journal}{Phys.Lett.} \textbf{\bibinfo{volume}{B78}},
  \bibinfo{pages}{150} (\bibinfo{year}{1978}).

\bibitem[{\citenamefont{Arsene et~al.}(2005)}]{Arsene:2004fa}
\bibinfo{author}{\bibfnamefont{I.}~\bibnamefont{Arsene}} \bibnamefont{et~al.}
  (\bibinfo{collaboration}{BRAHMS Collaboration}),
  \bibinfo{journal}{Nucl.Phys.} \textbf{\bibinfo{volume}{A757}},
  \bibinfo{pages}{1} (\bibinfo{year}{2005}), \eprint{nucl-ex/0410020}.

\bibitem[{\citenamefont{Back et~al.}(2005)}]{Back:2004je}
\bibinfo{author}{\bibfnamefont{B.}~\bibnamefont{Back}} \bibnamefont{et~al.}
  (\bibinfo{collaboration}{PHOBOS Collaboration}),
  \bibinfo{journal}{Nucl.Phys.} \textbf{\bibinfo{volume}{A757}},
  \bibinfo{pages}{28} (\bibinfo{year}{2005}), \eprint{nucl-ex/0410022}.

\bibitem[{\citenamefont{Adcox et~al.}(2005)}]{Adcox:2004mh}
\bibinfo{author}{\bibfnamefont{K.}~\bibnamefont{Adcox}} \bibnamefont{et~al.}
  (\bibinfo{collaboration}{PHENIX Collaboration}),
  \bibinfo{journal}{Nucl.Phys.} \textbf{\bibinfo{volume}{A757}},
  \bibinfo{pages}{184} (\bibinfo{year}{2005}), \eprint{nucl-ex/0410003}.

\bibitem[{\citenamefont{Adams et~al.}(2005{\natexlab{a}})}]{Adams:2005dq}
\bibinfo{author}{\bibfnamefont{J.}~\bibnamefont{Adams}} \bibnamefont{et~al.}
  (\bibinfo{collaboration}{STAR Collaboration}), \bibinfo{journal}{Nucl.Phys.}
  \textbf{\bibinfo{volume}{A757}}, \bibinfo{pages}{102}
  (\bibinfo{year}{2005}{\natexlab{a}}), \eprint{nucl-ex/0501009}.

\bibitem[{\citenamefont{Muller et~al.}(2012)\citenamefont{Muller, Schukraft,
  and Wyslouch}}]{Muller:2012zq}
\bibinfo{author}{\bibfnamefont{B.}~\bibnamefont{Muller}},
  \bibinfo{author}{\bibfnamefont{J.}~\bibnamefont{Schukraft}},
  \bibnamefont{and} \bibinfo{author}{\bibfnamefont{B.}~\bibnamefont{Wyslouch}},
  \bibinfo{journal}{Ann.Rev.Nucl.Part.Sci.} \textbf{\bibinfo{volume}{62}},
  \bibinfo{pages}{361} (\bibinfo{year}{2012}), \eprint{1202.3233}.

\bibitem[{\citenamefont{Adcox et~al.}(2002)}]{Adcox:2001jp}
\bibinfo{author}{\bibfnamefont{K.}~\bibnamefont{Adcox}} \bibnamefont{et~al.}
  (\bibinfo{collaboration}{PHENIX Collaboration}),
  \bibinfo{journal}{Phys.Rev.Lett.} \textbf{\bibinfo{volume}{88}},
  \bibinfo{pages}{022301} (\bibinfo{year}{2002}), \eprint{nucl-ex/0109003}.

\bibitem[{\citenamefont{Adler et~al.}(2002)}]{Adler:2002xw}
\bibinfo{author}{\bibfnamefont{C.}~\bibnamefont{Adler}} \bibnamefont{et~al.}
  (\bibinfo{collaboration}{STAR Collaboration}),
  \bibinfo{journal}{Phys.Rev.Lett.} \textbf{\bibinfo{volume}{89}},
  \bibinfo{pages}{202301} (\bibinfo{year}{2002}), \eprint{nucl-ex/0206011}.

\bibitem[{\citenamefont{Adler et~al.}(2003{\natexlab{a}})}]{Adler:2002tq}
\bibinfo{author}{\bibfnamefont{C.}~\bibnamefont{Adler}} \bibnamefont{et~al.}
  (\bibinfo{collaboration}{STAR Collaboration}),
  \bibinfo{journal}{Phys.Rev.Lett.} \textbf{\bibinfo{volume}{90}},
  \bibinfo{pages}{082302} (\bibinfo{year}{2003}{\natexlab{a}}),
  \eprint{nucl-ex/0210033}.

\bibitem[{\citenamefont{Adams et~al.}(2003{\natexlab{a}})}]{Adams:2003kv}
\bibinfo{author}{\bibfnamefont{J.}~\bibnamefont{Adams}} \bibnamefont{et~al.}
  (\bibinfo{collaboration}{STAR Collaboration}),
  \bibinfo{journal}{Phys.Rev.Lett.} \textbf{\bibinfo{volume}{91}},
  \bibinfo{pages}{172302} (\bibinfo{year}{2003}{\natexlab{a}}),
  \eprint{nucl-ex/0305015}.

\bibitem[{\citenamefont{Adler et~al.}(2003{\natexlab{b}})}]{Adler:2003ii}
\bibinfo{author}{\bibfnamefont{S.}~\bibnamefont{Adler}} \bibnamefont{et~al.}
  (\bibinfo{collaboration}{PHENIX Collaboration}),
  \bibinfo{journal}{Phys.Rev.Lett.} \textbf{\bibinfo{volume}{91}},
  \bibinfo{pages}{072303} (\bibinfo{year}{2003}{\natexlab{b}}),
  \eprint{nucl-ex/0306021}.

\bibitem[{\citenamefont{Adler et~al.}(2003{\natexlab{c}})}]{Adler:2003qi}
\bibinfo{author}{\bibfnamefont{S.}~\bibnamefont{Adler}} \bibnamefont{et~al.}
  (\bibinfo{collaboration}{PHENIX Collaboration}),
  \bibinfo{journal}{Phys.Rev.Lett.} \textbf{\bibinfo{volume}{91}},
  \bibinfo{pages}{072301} (\bibinfo{year}{2003}{\natexlab{c}}),
  \eprint{nucl-ex/0304022}.

\bibitem[{\citenamefont{Adams et~al.}(2003{\natexlab{b}})}]{Adams:2003im}
\bibinfo{author}{\bibfnamefont{J.}~\bibnamefont{Adams}} \bibnamefont{et~al.}
  (\bibinfo{collaboration}{STAR Collaboration}),
  \bibinfo{journal}{Phys.Rev.Lett.} \textbf{\bibinfo{volume}{91}},
  \bibinfo{pages}{072304} (\bibinfo{year}{2003}{\natexlab{b}}),
  \eprint{nucl-ex/0306024}.

%kjiang add3
\bibitem[{\citenamefont{Aad et~al.}(2010)}]{Aad:2010bu}
\bibinfo{author}{\bibfnamefont{G.}~\bibnamefont{Aad}} \bibnamefont{et~al.} (\bibinfo{collaboration}{ATLAS Collaboration}),
\bibinfo{journal}{Phys.Rev.Lett.} \textbf{\bibinfo{volume}{105}},
\bibinfo{pages}{252303} (\bibinfo{year}{2010}), 
\eprint{1011.6182}.

\bibitem[{\citenamefont{Chatrchyan et~al.}(2011)}]{Chatrchyan:2011sx}
\bibinfo{author}{\bibfnamefont{S.}~\bibnamefont{Chatrchyan}} \bibnamefont{et~al.} (\bibinfo{collaboration}{CMS Collaboration}), 
\bibinfo{journal}{Phys.Rev.} \textbf{\bibinfo{volume}{C84}}, 
\bibinfo{pages}{024906} (\bibinfo{year}{2011}), 
\eprint{1102.1957}.

\bibitem[{\citenamefont{Aamodt et~al.}(2011)}]{Aamodt:2010jd}
\bibinfo{author}{\bibfnamefont{K.}~\bibnamefont{Aamodt}} \bibnamefont{et~al.} (\bibinfo{collaboration}{ALICE Collaboration}), 
\bibinfo{journal}{Phys.Lett.} \textbf{\bibinfo{volume}{B696}},
\bibinfo{pages}{30} (\bibinfo{year}{2011}).
\eprint{1012.1004}.

\bibitem[{\citenamefont{Jacobs and Wang}(2005)}]{Jacobs:2004qv}
\bibinfo{author}{\bibfnamefont{P.}~\bibnamefont{Jacobs}} \bibnamefont{and}
  \bibinfo{author}{\bibfnamefont{X.-N.} \bibnamefont{Wang}},
  \bibinfo{journal}{Prog.Part.Nucl.Phys.} \textbf{\bibinfo{volume}{54}},
  \bibinfo{pages}{443} (\bibinfo{year}{2005}), \eprint{hep-ph/0405125}.

\bibitem[{\citenamefont{Stoecker}(2005)}]{Stoecker:2004qu}
\bibinfo{author}{\bibfnamefont{H.}~\bibnamefont{Stoecker}},
  \bibinfo{journal}{Nucl.Phys.} \textbf{\bibinfo{volume}{A750}},
  \bibinfo{pages}{121} (\bibinfo{year}{2005}), \eprint{nucl-th/0406018}.

\bibitem[{\citenamefont{Casalderrey-Solana
  et~al.}(2005)\citenamefont{Casalderrey-Solana, Shuryak, and
  Teaney}}]{CasalderreySolana:2004qm}
\bibinfo{author}{\bibfnamefont{J.}~\bibnamefont{Casalderrey-Solana}},
  \bibinfo{author}{\bibfnamefont{E.}~\bibnamefont{Shuryak}}, \bibnamefont{and}
  \bibinfo{author}{\bibfnamefont{D.}~\bibnamefont{Teaney}},
  \bibinfo{journal}{J.Phys.Conf.Ser.} \textbf{\bibinfo{volume}{27}},
  \bibinfo{pages}{22} (\bibinfo{year}{2005}), \eprint{hep-ph/0411315}.

%kjiang
\bibitem[{\citenamefont{Ma et~al.}(2011)\citenamefont{Ma, and Wang}}]{Ma:2010dv}
\bibinfo{author}{\bibfnamefont{G.-L.}~\bibnamefont{Ma}}, \bibnamefont{and}
\bibinfo{author}{\bibfnamefont{X.-N.}~\bibnamefont{Wang}},
\bibinfo{journal}{Phys.Rev.Lett.} \textbf{\bibinfo{volume}{106}},
\bibinfo{pages}{162301} (\bibinfo{year}{2011}), 
\eprint{1011.5249}.

\bibitem[{\citenamefont{Tachibana et~al.}(2014)\citenamefont{Tachibana, and Hirano}}]{Tachibana:2014lja}
\bibinfo{author}{\bibfnamefont{Y.}~\bibnamefont{Tachibana}}, \bibnamefont{and}
\bibinfo{author}{\bibfnamefont{T.}~\bibnamefont{Hirano}},
\bibinfo{journal}{Phys.Rev.} \textbf{\bibinfo{volume}{C90}}, 
\bibinfo{pages}{021902} (\bibinfo{year}{2014}), 
\eprint{1402.6469}.

\bibitem[{\citenamefont{Tachibana et~al.}(2016)\citenamefont{Tachibana, and Hirano}}]{Tachibana:2015qxa}
\bibinfo{author}{\bibfnamefont{Y.}~\bibnamefont{Tachibana}}, \bibnamefont{and}
\bibinfo{author}{\bibfnamefont{T.}~\bibnamefont{Hirano}},
\bibinfo{journal}{Phys.Rev.} \textbf{\bibinfo{volume}{C93}}, 
\bibinfo{pages}{054907} (\bibinfo{year}{2016}), 
\eprint{1510.06966}.

\bibitem[{\citenamefont{Tachibana et~al.}(2017)\citenamefont{Tachibana, Chang, and Qin}}]{Tachibana:2017syd}
\bibinfo{author}{\bibfnamefont{Y.}~\bibnamefont{Tachibana}},
\bibinfo{author}{\bibfnamefont{N.-B.}~\bibnamefont{Chang}}, \bibnamefont{and}
\bibinfo{author}{\bibfnamefont{G.-Y.}~\bibnamefont{Qin}},
\bibinfo{journal}{Phys.Rev.} \textbf{\bibinfo{volume}{C95}}, 
\bibinfo{pages}{044909} (\bibinfo{year}{2017}), 
\eprint{1701.07951}.

%kjiang
\bibitem[{\citenamefont{Adare et~al.}(2013)}]{Adare:2012qi}
\bibinfo{author}{\bibfnamefont{A.}~\bibnamefont{Adare}} \bibnamefont{et~al.} (\bibinfo{collaboration}{PHENIX Collaboration}),
\bibinfo{journal}{Phys.Rev.Lett.} \textbf{\bibinfo{volume}{111}},
\bibinfo{pages}{032301} (\bibinfo{year}{2013}), 
\eprint{1212.3323}.

\bibitem[{\citenamefont{Khachatryan et~al.}(2016)}]{Khachatryan:2016erx}
\bibinfo{author}{\bibfnamefont{V.}~\bibnamefont{Khachatryan}} \bibnamefont{et~al.} (\bibinfo{collaboration}{CMS Collaboration}),
\bibinfo{journal}{JHEP} \textbf{\bibinfo{volume}{02}},
\bibinfo{pages}{156} (\bibinfo{year}{2016}), 
\eprint{1601.00079}.

\bibitem[{\citenamefont{Chatrchyan et~al.}(2014)}]{Chatrchyan:2013kwa}
\bibinfo{author}{\bibfnamefont{S.}~\bibnamefont{Chatrchyan}} \bibnamefont{et~al.} (\bibinfo{collaboration}{CMS Collaboration}), 
\bibinfo{journal}{Phys.Lett.} \textbf{\bibinfo{volume}{B730}},
\bibinfo{pages}{243} (\bibinfo{year}{2014}).
\eprint{1310.0878}.



\bibitem[{\citenamefont{Adams et~al.}(2005{\natexlab{b}})}]{Adams:2005ph}
\bibinfo{author}{\bibfnamefont{J.}~\bibnamefont{Adams}} \bibnamefont{et~al.}
  (\bibinfo{collaboration}{STAR Collaboration}),
  \bibinfo{journal}{Phys.Rev.Lett.} \textbf{\bibinfo{volume}{95}},
  \bibinfo{pages}{152301} (\bibinfo{year}{2005}{\natexlab{b}}),
  \eprint{nucl-ex/0501016}.

\bibitem[{\citenamefont{Adler et~al.}(2006{\natexlab{a}})}]{Adler:2005ee}
\bibinfo{author}{\bibfnamefont{S.}~\bibnamefont{Adler}} \bibnamefont{et~al.}
  (\bibinfo{collaboration}{PHENIX Collaboration}),
  \bibinfo{journal}{Phys.Rev.Lett.} \textbf{\bibinfo{volume}{97}},
  \bibinfo{pages}{052301} (\bibinfo{year}{2006}{\natexlab{a}}),
  \eprint{nucl-ex/0507004}.

\bibitem[{\citenamefont{Adare et~al.}(2008)}]{Adare:2008ae}
\bibinfo{author}{\bibfnamefont{A.}~\bibnamefont{Adare}} \bibnamefont{et~al.}
  (\bibinfo{collaboration}{PHENIX Collaboration}), \bibinfo{journal}{Phys.Rev.}
  \textbf{\bibinfo{volume}{C78}}, \bibinfo{pages}{014901}
  (\bibinfo{year}{2008}), \eprint{0801.4545}.

\bibitem[{\citenamefont{Aggarwal et~al.}(2010)}]{Aggarwal:2010rf}
\bibinfo{author}{\bibfnamefont{M.}~\bibnamefont{Aggarwal}} \bibnamefont{et~al.}
  (\bibinfo{collaboration}{STAR Collaboration}), \bibinfo{journal}{Phys.Rev.}
  \textbf{\bibinfo{volume}{C82}}, \bibinfo{pages}{024912}
  (\bibinfo{year}{2010}), \eprint{1004.2377}.

\bibitem[{\citenamefont{Wang}(2014)}]{Wang:2013qca}
\bibinfo{author}{\bibfnamefont{F.}~\bibnamefont{Wang}}, \bibinfo{journal}{Prog.
  Part. Nucl. Phys.} \textbf{\bibinfo{volume}{74}}, \bibinfo{pages}{35}
  (\bibinfo{year}{2014}), \eprint{1311.4444}.

\bibitem[{\citenamefont{Alver and Roland}(2010)}]{Alver:2010gr}
\bibinfo{author}{\bibfnamefont{B.}~\bibnamefont{Alver}} \bibnamefont{and}
  \bibinfo{author}{\bibfnamefont{G.}~\bibnamefont{Roland}},
  \bibinfo{journal}{Phys.Rev.} \textbf{\bibinfo{volume}{C81}},
  \bibinfo{pages}{054905} (\bibinfo{year}{2010}),
  \bibinfo{note}{erratum-ibid.~{\bf C82}, 039903 (2010)}, \eprint{1003.0194}.

\bibitem[{\citenamefont{Heinz and Snellings}(2013)}]{Heinz:2013th}
\bibinfo{author}{\bibfnamefont{U.}~\bibnamefont{Heinz}} \bibnamefont{and}
  \bibinfo{author}{\bibfnamefont{R.}~\bibnamefont{Snellings}},
  \bibinfo{journal}{Ann.Rev.Nucl.Part.Sci.} \textbf{\bibinfo{volume}{63}},
  \bibinfo{pages}{123} (\bibinfo{year}{2013}), \eprint{1301.2826}.

\bibitem[{\citenamefont{Agakishiev et~al.}(2010)}]{Agakishiev:2010ur}
\bibinfo{author}{\bibfnamefont{H.}~\bibnamefont{Agakishiev}}
  \bibnamefont{et~al.} (\bibinfo{collaboration}{STAR Collaboration})
  (\bibinfo{year}{2010}), \eprint{1010.0690}.

\bibitem[{\citenamefont{Agakishiev et~al.}(2014)}]{Agakishiev:2014ada}
\bibinfo{author}{\bibfnamefont{H.}~\bibnamefont{Agakishiev}}
  \bibnamefont{et~al.} (\bibinfo{collaboration}{STAR}), \bibinfo{journal}{Phys.
  Rev.} \textbf{\bibinfo{volume}{C89}}, \bibinfo{pages}{041901}
  (\bibinfo{year}{2014}), \eprint{1404.1070}.

\bibitem[{\citenamefont{Adare et~al.}(2019)}]{Adare:2018wjb}
\bibinfo{author}{\bibfnamefont{A.}~\bibnamefont{Adare}} \bibnamefont{et~al.}
  (\bibinfo{collaboration}{PHENIX}), \bibinfo{journal}{Phys. Rev.}
  \textbf{\bibinfo{volume}{C99}}, \bibinfo{pages}{054903}
  (\bibinfo{year}{2019}), \eprint{1803.01749}.

%kjiang
\bibitem[{\citenamefont{Aamodt et~al.}(2012)}]{Aamodt:2011vg}
\bibinfo{author}{\bibfnamefont{K.}~\bibnamefont{Aamodt}} \bibnamefont{et~al.} (\bibinfo{collaboration}{ALICE Collaboration}),
\bibinfo{journal}{Phys.Rev.Lett.} \textbf{\bibinfo{volume}{108}},
\bibinfo{pages}{092301} (\bibinfo{year}{2012}), 
\eprint{1110.0121}.

\bibitem[{\citenamefont{Adam et~al.}(2016)}]{Adam:2016xbp}
\bibinfo{author}{\bibfnamefont{J.}~\bibnamefont{Adam}} \bibnamefont{et~al.} (\bibinfo{collaboration}{ALICE Collaboration}), 
\bibinfo{journal}{Phys.Lett.} \textbf{\bibinfo{volume}{B763}},
\bibinfo{pages}{238} (\bibinfo{year}{2016}).
\eprint{1608.07201}.

\bibitem[{\citenamefont{Adamczyk et~al.}(2016)}]{STAR:2016jdz}
\bibinfo{author}{\bibfnamefont{L.}~\bibnamefont{Adamczyk}} \bibnamefont{et~al.} (\bibinfo{collaboration}{STAR Collaboration}), 
\bibinfo{journal}{Phys.Lett.} \textbf{\bibinfo{volume}{B760}},
\bibinfo{pages}{689} (\bibinfo{year}{2016}).
\eprint{1604.01117}.

\bibitem[{\citenamefont{Adamczyk et~al.}(2014)}]{Adamczyk:2013jei}
\bibinfo{author}{\bibfnamefont{L.}~\bibnamefont{Adamczyk}} \bibnamefont{et~al.} (\bibinfo{collaboration}{STAR Collaboration}),
\bibinfo{journal}{Phys.Rev.Lett.} \textbf{\bibinfo{volume}{112}},
\bibinfo{pages}{122301} (\bibinfo{year}{2014}), 
\eprint{1302.6184}.

\bibitem[{\citenamefont{Adamczyk
  et~al.}(2013{\natexlab{a}})}]{Adamczyk:2012eoa}
\bibinfo{author}{\bibfnamefont{L.}~\bibnamefont{Adamczyk}} \bibnamefont{et~al.}
  (\bibinfo{collaboration}{STAR}), \bibinfo{journal}{Phys. Rev.}
  \textbf{\bibinfo{volume}{C87}}, \bibinfo{pages}{044903}
  (\bibinfo{year}{2013}{\natexlab{a}}), \eprint{1212.1653}.

\bibitem[{\citenamefont{Agakishiev et~al.}(2012)}]{Agakishiev:2011st}
\bibinfo{author}{\bibfnamefont{G.}~\bibnamefont{Agakishiev}} \bibnamefont{et~al.} (\bibinfo{collaboration}{STAR Collaboration}), 
\bibinfo{journal}{Phys.Rev.} \textbf{\bibinfo{volume}{C85}}, 
\bibinfo{pages}{014903} (\bibinfo{year}{2012}), 
\eprint{1110.5800}.

\bibitem[{\citenamefont{Abelev et~al.}(2016)}]{Abelev:2016dqe}
\bibinfo{author}{\bibfnamefont{B.}~\bibnamefont{Abelev}} \bibnamefont{et~al.} (\bibinfo{collaboration}{STAR Collaboration}), 
\bibinfo{journal}{Phys.Rev.} \textbf{\bibinfo{volume}{C94}}, 
\bibinfo{pages}{014910} (\bibinfo{year}{2016}), 
\eprint{1603.05477}.


\bibitem[{\citenamefont{Zhang et~al.}(2019)\citenamefont{Zhang, Jiang, Li, Liu, and Wang}}]{Zhang:2019lmn}
\bibinfo{author}{\bibfnamefont{L.}~\bibnamefont{Zhang}},
  \bibinfo{author}{\bibfnamefont{K.}~\bibnamefont{Jiang}},
  \bibinfo{author}{\bibfnamefont{C.}~\bibnamefont{Li}},
  \bibinfo{author}{\bibfnamefont{F.}~\bibnamefont{Liu}}, \bibnamefont{and}
  \bibinfo{author}{\bibfnamefont{F.}~\bibnamefont{Wang}}
\bibinfo{journal}{Phys.Rev.} \textbf{\bibinfo{volume}{C100}}, 
\bibinfo{pages}{014903} (\bibinfo{year}{2019}), 
  \eprint{1902.06027}.





\bibitem[{\citenamefont{Llope et~al.}(2004)\citenamefont{Llope, Geurts,
  Mitchell, Liu, Adams et~al.}}]{Llope:2003ti}
\bibinfo{author}{\bibfnamefont{W.}~\bibnamefont{Llope}},
  \bibinfo{author}{\bibfnamefont{F.}~\bibnamefont{Geurts}},
  \bibinfo{author}{\bibfnamefont{J.}~\bibnamefont{Mitchell}},
  \bibinfo{author}{\bibfnamefont{Z.}~\bibnamefont{Liu}},
  \bibinfo{author}{\bibfnamefont{N.}~\bibnamefont{Adams}},
  \bibnamefont{et~al.}, \bibinfo{journal}{Nucl.Instrum.Meth.}
  \textbf{\bibinfo{volume}{A522}}, \bibinfo{pages}{252} (\bibinfo{year}{2004}),
  \eprint{nucl-ex/0308022}.

\bibitem[{\citenamefont{Abelev et~al.}(2009{\natexlab{a}})}]{Abelev:2008ab}
\bibinfo{author}{\bibfnamefont{B.}~\bibnamefont{Abelev}} \bibnamefont{et~al.}
  (\bibinfo{collaboration}{STAR Collaboration}), \bibinfo{journal}{Phys.Rev.}
  \textbf{\bibinfo{volume}{C79}}, \bibinfo{pages}{034909}
  (\bibinfo{year}{2009}{\natexlab{a}}), \eprint{0808.2041}.

\bibitem[{\citenamefont{Ackermann et~al.}(1999)}]{Ackermann:1999kc}
\bibinfo{author}{\bibfnamefont{K.}~\bibnamefont{Ackermann}}
  \bibnamefont{et~al.} (\bibinfo{collaboration}{STAR Collaboration}),
  \bibinfo{journal}{Nucl.Phys.} \textbf{\bibinfo{volume}{A661}},
  \bibinfo{pages}{681} (\bibinfo{year}{1999}).

\bibitem[{\citenamefont{Anderson et~al.}(2003)}]{Anderson:2003ur}
\bibinfo{author}{\bibfnamefont{M.}~\bibnamefont{Anderson}}
  \bibnamefont{et~al.}, \bibinfo{journal}{Nucl.Instrum.Meth.}
  \textbf{\bibinfo{volume}{A499}}, \bibinfo{pages}{659} (\bibinfo{year}{2003}),
  \eprint{nucl-ex/0301015}.

\bibitem[{\citenamefont{Zhang et~al.}(2007)\citenamefont{Zhang, Owens, Wang,
  and Wang}}]{Zhang:2007ja}
\bibinfo{author}{\bibfnamefont{H.}~\bibnamefont{Zhang}},
  \bibinfo{author}{\bibfnamefont{J.~F.} \bibnamefont{Owens}},
  \bibinfo{author}{\bibfnamefont{E.}~\bibnamefont{Wang}}, \bibnamefont{and}
  \bibinfo{author}{\bibfnamefont{X.-N.} \bibnamefont{Wang}},
  \bibinfo{journal}{Phys. Rev. Lett.} \textbf{\bibinfo{volume}{98}},
  \bibinfo{pages}{212301} (\bibinfo{year}{2007}), \eprint{nucl-th/0701045}.


\bibitem[{\citenamefont{Renk}(2013)}]{Renk:2012ve}
\bibinfo{author}{\bibfnamefont{T.}~\bibnamefont{Renk}}, 
\bibinfo{journal}{Phys.Rev.} \textbf{\bibinfo{volume}{C88}}, 
\bibinfo{pages}{054902} (\bibinfo{year}{2013}), 
\eprint{1212.0646}.

\bibitem[{\citenamefont{Agakishiev et~al.}(2011)}]{Agakishiev:2011nb}
\bibinfo{author}{\bibfnamefont{H.}~\bibnamefont{Agakishiev}}
  \bibnamefont{et~al.} (\bibinfo{collaboration}{STAR}), \bibinfo{journal}{Phys.
  Rev.} \textbf{\bibinfo{volume}{C83}}, \bibinfo{pages}{061901}
  (\bibinfo{year}{2011}), \eprint{1102.2669}.


\bibitem[{\citenamefont{Bozek et~al.}(2011)\citenamefont{Bozek, Broniowski, and
  Moreira}}]{Bozek:2010vz}
\bibinfo{author}{\bibfnamefont{P.}~\bibnamefont{Bozek}},
  \bibinfo{author}{\bibfnamefont{W.}~\bibnamefont{Broniowski}},
  \bibnamefont{and} \bibinfo{author}{\bibfnamefont{J.}~\bibnamefont{Moreira}},
  \bibinfo{journal}{Phys.Rev.} \textbf{\bibinfo{volume}{C83}},
  \bibinfo{pages}{034911} (\bibinfo{year}{2011}), \eprint{1011.3354}.

\bibitem[{\citenamefont{Xiao et~al.}(2013)\citenamefont{Xiao, Liu, and
  Wang}}]{Xiao:2012uw}
\bibinfo{author}{\bibfnamefont{K.}~\bibnamefont{Xiao}},
  \bibinfo{author}{\bibfnamefont{F.}~\bibnamefont{Liu}}, \bibnamefont{and}
  \bibinfo{author}{\bibfnamefont{F.}~\bibnamefont{Wang}},
  \bibinfo{journal}{Phys.Rev.} \textbf{\bibinfo{volume}{C87}},
  \bibinfo{pages}{011901} (\bibinfo{year}{2013}), \eprint{1208.1195}.

\bibitem[{\citenamefont{Jia and Huo}(2014)}]{Jia:2014ysa}
\bibinfo{author}{\bibfnamefont{J.}~\bibnamefont{Jia}} \bibnamefont{and}
  \bibinfo{author}{\bibfnamefont{P.}~\bibnamefont{Huo}},
  \bibinfo{journal}{Phys.Rev.} \textbf{\bibinfo{volume}{C90}},
  \bibinfo{pages}{034915} (\bibinfo{year}{2014}), \eprint{1403.6077}.

\bibitem[{\citenamefont{Khachatryan et~al.}(2015)}]{Khachatryan:2015oea}
\bibinfo{author}{\bibfnamefont{V.}~\bibnamefont{Khachatryan}}
  \bibnamefont{et~al.} (\bibinfo{collaboration}{CMS}), \bibinfo{journal}{Phys.
  Rev.} \textbf{\bibinfo{volume}{C92}}, \bibinfo{pages}{034911}
  (\bibinfo{year}{2015}), \eprint{1503.01692}.

\bibitem[{\citenamefont{Cronin et~al.}(1975)\citenamefont{Cronin, Frisch,
  Shochet, Boymond, Mermod, Piroue, and Sumner}}]{Cronin:1974zm}
\bibinfo{author}{\bibfnamefont{J.~W.} \bibnamefont{Cronin}},
  \bibinfo{author}{\bibfnamefont{H.~J.} \bibnamefont{Frisch}},
  \bibinfo{author}{\bibfnamefont{M.~J.} \bibnamefont{Shochet}},
  \bibinfo{author}{\bibfnamefont{J.~P.} \bibnamefont{Boymond}},
  \bibinfo{author}{\bibfnamefont{R.}~\bibnamefont{Mermod}},
  \bibinfo{author}{\bibfnamefont{P.~A.} \bibnamefont{Piroue}},
  \bibnamefont{and} \bibinfo{author}{\bibfnamefont{R.~L.}
  \bibnamefont{Sumner}}, \bibinfo{journal}{Phys. Rev.}
  \textbf{\bibinfo{volume}{D11}}, \bibinfo{pages}{3105} (\bibinfo{year}{1975}).

\bibitem[{\citenamefont{Vitev}(2005)}]{Vitev:2004kd}
\bibinfo{author}{\bibfnamefont{I.}~\bibnamefont{Vitev}}, \bibinfo{journal}{J.
  Phys.} \textbf{\bibinfo{volume}{G31}}, \bibinfo{pages}{S557}
  (\bibinfo{year}{2005}), \eprint{hep-ph/0409297}.

\bibitem[{\citenamefont{Rak}(2004)}]{Rak:2004gk}
\bibinfo{author}{\bibfnamefont{J.}~\bibnamefont{Rak}}, \bibinfo{journal}{J.
  Phys.} \textbf{\bibinfo{volume}{G30}}, \bibinfo{pages}{S1309}
  (\bibinfo{year}{2004}), \eprint{hep-ex/0403038}.

\bibitem[{\citenamefont{Adler et~al.}(2006{\natexlab{b}})}]{Adler:2006sc}
\bibinfo{author}{\bibfnamefont{S.}~\bibnamefont{Adler}} \bibnamefont{et~al.}
  (\bibinfo{collaboration}{PHENIX Collaboration}), \bibinfo{journal}{Phys.Rev.}
  \textbf{\bibinfo{volume}{D74}}, \bibinfo{pages}{072002}
  (\bibinfo{year}{2006}{\natexlab{b}}), \eprint{hep-ex/0605039}.

\bibitem[{\citenamefont{Henry}(2006)}]{Henry:2005wr}
\bibinfo{author}{\bibfnamefont{T.}~\bibnamefont{Henry}}
  (\bibinfo{collaboration}{STAR}), \bibinfo{journal}{Acta Phys. Hung.}
  \textbf{\bibinfo{volume}{A27}}, \bibinfo{pages}{217} (\bibinfo{year}{2006}),
  \eprint{nucl-ex/0511002}.

\bibitem[{\citenamefont{Putschke}(2009)}]{Putschke:2009wr}
\bibinfo{author}{\bibfnamefont{J.}~\bibnamefont{Putschke}}
  (\bibinfo{collaboration}{STAR}), \bibinfo{journal}{Nucl. Phys.}
  \textbf{\bibinfo{volume}{A830}}, \bibinfo{pages}{58C} (\bibinfo{year}{2009}),
  \eprint{0908.1766}.

\bibitem[{\citenamefont{Abelev et~al.}(2009{\natexlab{b}})}]{Abelev:2008ac}
\bibinfo{author}{\bibfnamefont{B.}~\bibnamefont{Abelev}} \bibnamefont{et~al.}
  (\bibinfo{collaboration}{STAR Collaboration}),
  \bibinfo{journal}{Phys.Rev.Lett.} \textbf{\bibinfo{volume}{102}},
  \bibinfo{pages}{052302} (\bibinfo{year}{2009}{\natexlab{b}}),
  \eprint{0805.0622}.

\bibitem[{\citenamefont{Abelev et~al.}(2010)}]{Abelev:2009jv}
\bibinfo{author}{\bibfnamefont{B.}~\bibnamefont{Abelev}} \bibnamefont{et~al.}
  (\bibinfo{collaboration}{STAR Collaboration}),
  \bibinfo{journal}{Phys.Rev.Lett.} \textbf{\bibinfo{volume}{105}},
  \bibinfo{pages}{022301} (\bibinfo{year}{2010}), \eprint{0912.3977}.

\end{thebibliography}
